\input harvmac.tex
\overfullrule=0mm

\input epsf.tex
\newcount\figno
\figno=0
\def\fig#1#2#3{
\par\begingroup\parindent=0pt\leftskip=1cm\rightskip=1cm\parindent=0pt
\baselineskip=11pt
\global\advance\figno by 1
\midinsert
\epsfxsize=#3
\centerline{\epsfbox{#2}}
\vskip 12pt
{\bf Fig. \the\figno:} #1\par
\endinsert\endgroup\par
}
\def\figlabel#1{\xdef#1{\the\figno}}
\def\encadremath#1{\vbox{\hrule\hbox{\vrule\kern8pt\vbox{\kern8pt
\hbox{$\displaystyle #1$}\kern8pt}
\kern8pt\vrule}\hrule}}


%
\def\frac#1#2{\scriptstyle{#1 \over #2}}

\def\ket#1{ | #1 \rangle}
\def\bra#1{ \langle #1 |}
\def\tp{t_{{\textstyle .}}}
%
%
\def\CA{{\cal A}}		
\def\CD{{\cal D}}	\def\CE{{\cal E}}	
\def\CG{{\cal G}}	\def\CH{{\cal H}}	\def\CI{{\cal J}}
	\def\CK{{\cal K}}	
		\def\CO{{\cal O}}
\def\CP{{\cal P}}		
	\def\CT{{\cal T}}	\def\CU{{\cal U}}

\def\({ \left( }\def\[{ \left[ }
\def\){ \right) }\def\]{ \right] }
%


\def\IR{\relax{\rm I\kern-.18em R}}
\font\cmss=cmss10 \font\cmsss=cmss10 at 7pt
\def\IZ{\relax\ifmmode\mathchoice
{\hbox{\cmss Z\kern-.4em Z}}{\hbox{\cmss Z\kern-.4em Z}}
{\lower.9pt\hbox{\cmsss Z\kern-.4em Z}}
{\lower1.2pt\hbox{\cmsss Z\kern-.4em Z}}\else{\cmss Z\kern-.4em Z}\fi}
\def\inbar{\,\vrule height1.5ex width.4pt depth0pt}
\def\IB{\relax{\rm I\kern-.18em B}}
\def\IC{\relax\hbox{$\inbar\kern-.3em{\rm C}$}}
\def\ID{\relax{\rm I\kern-.18em D}}
\def\IE{\relax{\rm I\kern-.18em E}}
\def\IF{\relax{\rm I\kern-.18em F}}
\def\IG{\relax\hbox{$\inbar\kern-.3em{\rm G}$}}
\def\IH{\relax{\rm I\kern-.18em H}}
\def\II{\relax{\rm I\kern-.18em I}}
\def\IK{\relax{\rm I\kern-.18em K}}
\def\IL{\relax{\rm I\kern-.18em L}}
\def\IM{\relax{\rm I\kern-.18em M}}
\def\IN{\relax{\rm I\kern-.18em N}}
\def\IO{\relax\hbox{$\inbar\kern-.3em{\rm O}$}}
\def\IP{\relax{\rm I\kern-.18em P}}
\def\IQ{\relax\hbox{$\inbar\kern-.3em{\rm Q}$}}
\def\IGa{\relax\hbox{${\rm I}\kern-.18em\Gamma$}}
\def\IPi{\relax\hbox{${\rm I}\kern-.18em\Pi$}}
\def\ITh{\relax\hbox{$\inbar\kern-.3em\Theta$}}
\def\IOm{\relax\hbox{$\inbar\kern-3.00pt\Omega$}}


\def\d{{\rm d}}

\def\oh{{1\over 2}}\def\un{{\bf 1}}

\def\Ga{\alpha}\def\Gc{\gamma}
\def\Gd{\delta}\def\Ge{\epsilon}

\def\Gl{\lambda}\def\GL{\Lambda}


\def\dim{{\rm dim\,}}\def\mod{{\rm mod\,}}

\def\bra{\langle}\def\ket{\rangle}
\def\nind{\par\noindent}
\def\td{t_{\textstyle{.}}}


\def\bi{\bar\imath}\def\bj{\bar\jmath}
\def\Exp{{\rm Exp}\,}  

%
\def\msy{y }
\message{Do you have the AMS fonts (y/n) ?}\read-1 to \msan
\ifx\msan\msy
\input amssym.def
\input amssym.tex
\def\IZ{\Bbb Z}\def\IR{\Bbb R}\def\IC{\Bbb C}\def\IN{\Bbb N}
\def\gg{\goth g}
\else 
\def\gg{g}
\fi
\Title{\vbox{\hbox{SPhT 94/156}
\hbox{{\tt hep-th/9412202   }}}}
{{\vbox {
\centerline{Conformal, Integrable and Topological Theories, }
\medskip
\centerline{Graphs and Coxeter Groups} 
}}}

\bigskip

\centerline{J.-B. Zuber}\bigskip

\centerline{ \it CEA, Service de Physique Th\'eorique de Saclay
,}
\centerline{ \it F-91191 Gif sur Yvette Cedex, France}

\vskip .2in

\noindent 
I review three different problems occuring in two dimensional field
theory: 1) classification of conformal field theories; 2) 
construction of lattice integrable  realizations of the latter; 
3) solutions to the WDVV equations of topological field theories. 
I show that a structure of Coxeter group is hidden behind these 
three related problems.

\Date{
12/94\quad
Talk delivered at the International Conference
of Mathematical Physics, Paris 18-23 July 1994 }
%



\lref\BPZ{A.A. Belavin, A.M. Polyakov and A.B. Zamolodchikov,
{\it  Nucl. Phys.} {\bf B241} (1984) 333.}
\lref\MoSe{ G. Moore and N. Seiberg, {\it Comm. Math. Phys. }{\bf 123} (1989)
177;
{\it Lectures on RCFT},
in {\it Superstrings {\oldstyle 89}}, 
proceedings of the 1989 Trieste spring school, M. Green {\it et al.} eds,
World Scientific 1990  and further references therein. }
\lref\JLC{J.L. Cardy, {\it Nucl. Phys.} {\bf B270} 
[FS16] (1986) 186.} 
\lref\WN{W. Nahm, {\it Nucl. Phys.} {\bf B114} (1976) 174.}
\lref\ABF{G.E. Andrews, R.J. Baxter and P.J. Forrester, {\it J.Stat. Phys.}
 {\bf 35} (1984) 193.}
\lref\VPun{V. Pasquier, {\it Nucl. Phys.} {\bf B285} [FS19] (1987) 162.} 
\lref\TL{P. Martin, {\it Potts models and related problems in statistical 
mechanics}, World Scientific, 1991
\semi {\it Yang Baxter Equations in Integrable Systems}, M. Jimbo edr, 
World Sc. 1989.}
\lref\Wi{E. Witten, Nucl. Phys. {\bf B 340} (1990) 281.}
\lref\DVV{R. Dijkgraaf, E. Verlinde and H. Verlinde, Nucl. Phys. 
{\bf B352} 59-86 (1991):
in {\it String Theory and Quantum Gravity},
proceedings of the 11990 Trieste Spring School, M. Green et al. {\it eds.},
World Sc. 1991.}
\lref\Dub{B. Dubrovin, Nucl. Phys. {\bf B 379} 627-689 (1992); 
{\it Differential Geometry of the space of orbits of a Coxeter group},
{\tt hep-th/9303152};
{\it Geometry of 2D Topological Field Theories}, {\tt hep-th/9407018}.}
\lref\Wa{N. Warner, {\it $N=2$ Supersymmetric Integrable Models and
Topological Field Theories}, to appear in the proceedings of the 1992 
Trieste Summer School, {\tt hep-th/9301088} and further references therein. }
\lref\GKO{P. Goddard, A. Kent and D. Olive, {\it Comm. Math. Phys. }
{\bf 103} (1986) 105 .}
\lref\KS{Y. Kazama and H. Suzuki, {\it Phys. Lett.} {\bf B216} 112 (1989); 
{\it Nucl. Phys.} {\bf B321} 232 (1989). }
\lref\CIZ{A. Cappelli, C. Itzykson, and J.-B. Zuber,
{\it Nucl. Phys.} {\bf B280} (1987) [FS]
445; {\it Comm. Math. Phys.} {\bf 113} (1987) 1
\semi A. Kato, {\it Mod. Phys. Lett.} {\bf A2} (1987) 585.}
\lref\DFZa{Vl.S. Dotsenko and V.A. Fateev,
{\it Nucl. Phys.} {\bf B240} (1984)
[FS] 312; {\it Nucl. Phys.} {\bf B251} (1985) [FS] 691;
{\it Phys. Lett.} {\bf154B} (1985) 291 \semi
A.B. Zamolodchikov and V.A. Fateev, {\it Sov. Phys. JETP} {\bf 62}
(1985) 215; {\it Sov. J. Nucl.Phys.} {\bf 43} (1986) 657 \semi
V.B. Petkova, {\it Int. J. Mod. Phys.} {\bf A3} (1988) 2945; 
V.B. Petkova, {\it Phys. Lett.} {\bf 225B} (1989) 357; 
P. Furlan, A.Ch. Ganchev and V.B. Petkova,
{\it Int. J. Mod. Phys.} {\bf A5} (1990) 2721; 
Erratum,  {\it ibid.} 3641 \semi
A. Kato and Y. Kitazawa, {\it Nucl. Phys.} {\bf B319} (1989) 474 \semi
J. Fuchs, {\it Phys. Rev. Lett.} {\bf 62} (1989) 1705; 
J. Fuchs and A. Klemm, {\it Ann.Phys.} (N.Y.) {\bf 194} (1989) 303; 
 J. Fuchs, {\it Phys. Lett.} {\bf 222B} (1989) 411; 
J. Fuchs, A. Klemm und C. Scheich, {\it Z. Phys.C} {\bf 46} (1990) 71\semi
M. Douglas and S. Trivedi, 
{\it Nucl. Phys.} {\bf B320} (1989) 461.}

\lref\PZ{V. Petkova and J.-B. Zuber, {\it On Structure Constants of $sl(2)$ Theories}, 
preprint ASI-TPA/20/94, SPhT 94/113, {\tt hep-th/9410209}. }
\lref\VP{V. Pasquier, J.Phys. {\bf A20} 5707-5717 (1987)} 
\lref\Gep{D. Gepner, {\it Nucl. Phys.} {\bf B296} (1988) 757; 
{\it Phys. Lett.} {\bf 199B } (1987) 380.} 
\lref\TN{T. Nassar, Rapport de DEA, Paris 1994.}
\lref\MVW{E. Martinec, {\it Phys. Lett.} 
{\bf B217} 431- (1989); {\it Criticality, 
catastrophes and compactifications}, in 
{\it Physics and mathematics of strings}, V.G. Knizhnik memorial volume,
L. Brink, D. Friedan and A.M. Polyakov eds., World Scientific 1990
\semi C. Vafa and N.P. Warner, {\it Phys. Lett.} {\bf B218} (1989) 51;
W. Lerche, C. Vafa, N.P. Warner, {\it Nucl. Phys.} {\bf B324} (1989) 427.}
\lref\AGZV{V.I. Arnold,  and S.M. Gusein-Zade, A.N. Varchenko,
{\it Singularities of differentiable maps}, Birk\"auser, Basel 1985.}
\lref\EYY{T. Eguchi, Y. Yamada and S.-K. Yang, 
{\it On the genus expansion in topological string theory},
{\tt hep-th 9405106}.}
\lref\BALZ{D. Bernard, {\it Nucl. Phys.}  {\bf B288} (1987) 628 \semi
D. Altsch\"uler, J. Lacki and P. Zaugg, {\it Phys. Lett.} 
{\bf 205B} (1988) 281 \semi
P. Christe and F. Ravanini, {\it Int. J. Mod. Phys.} {\bf A4} (1989) 897 \semi
G. Moore and N. Seiberg, {\it Nucl. Phys.} {\bf B313} (1989) 16 \semi
M. Bauer and C. Itzykson, {\it Comm. Math. Phys.} {\bf 127} (1990) 617 \semi
Ph. Ruelle, E. Thiran and J. Weyers, 
{\it Comm. Math. Phys. }{\bf 133} (1990) 305; 
{\it Nucl. Phys.} {\bf B402} (1993) 693.}
\lref\Gan{ T. Gannon, 
{\it Comm. Math. Phys.} {\bf 161} (1994) 233; 
{\it The classification of SU(3) modular invariants revisited},  
{\tt hep-th 9404185}. }
\lref\Bax{R. Baxter, {\it J. Stat. Phys.}, {\bf 28} (1982) 1.}
\lref\JMO{M. Jimbo, T. Miwa and M. Okado, {\it Lett. Math. Phys.} 
{\bf 14} (1987) 123; 
{\it Comm. Math. Phys.} {\bf 116} (1988) 507. } 
\lref\We{H. Wenzl, 
{\it Inv. Math.} {\bf 92} (1988) 349.}
\lref\Ko{I. Kostov, {\it Nucl. Phys.} {\bf B300} [FS22] (1988) 559.}
\lref\DFZ{P. Di Francesco and J.-B. Zuber, 
in {\it Recent Developments in Conformal Field Theories}, Trieste
Conference, 1989, S. Randjbar-Daemi, E. Sezgin and J.-B. Zuber eds., 
World Scientific 1990
\semi J.-B. Zuber, {\it Nucl. Phys.} (Proc. Suppl.) {\bf 18B} (1990) 313 
\semi P. Di Francesco, {\it Int.J.Mod.Phys.} {\bf A7} 407-500 (1992).}
\lref\So{N. Sochen, {\it Nucl. Phys.} {\bf B360} (1991) 613. }
\lref\DFLZ{
D. Nemeschansky, and N.P. Warner, Nucl. Phys. {\bf B380} 241-264 
(1992) 
\semi 
P. Di Francesco, F. Lesage and J.-B. Zuber,
Nucl. Phys. {\bf B408} (1993) 600.}
\lref\Zub{J.-B. Zuber, {\it Mod. Phys. Lett. A} {\bf 8} (1994) 749. }


\secno=-1
\newsec{Introduction}
\nind In this talk, I want to discuss connections between several
topics of current interest in mathematical physics, namely certain classes of
conformal field theories, integrable lattice models and topological 
field theories. This list is  not limitative, and might have included 
more items, such as integrable hierarchies and matrix models. 
The relations between these subjects 
are not very tight, certainly not one-to-one, 
 but they look strong enough to teach us about one starting 
from the other. 
The issue is first the classification programme of these various 
kinds of theories. But maybe even more important 
is to understand the principles that are beyond these classifications, and
the algebraic or geometric features common to these problems. 
As may be anticipated from the title of this contribution, 
I believe that such a common feature is to be found in the 
existence of a graph (or a collection of graphs); on the one hand,  
this graph, through 
its adjacency matrix and its eigenspectrum, codes some important data 
on the conformal theory or allows the construction of an integrable
model; on the other, it also codes the 
geometry of a root system and hence of a reflection (or Coxeter) group, 
that appears naturally in the study of 
topological theories, as shown recently by Dubrovin. 
Although I do not have yet a complete and consistent picture, 
the evidence seems compelling enough. 
I have tried to make this contribution 
readable by non specialists. Thus the first three sections
will be made of material that can  hardly be called original, 
namely a lightning review of the three 
kinds of problems that we are interested in and  the discussion of the 
results in the cases of theories based on $SU(2)$ and of $SU(3)$. 
Only in sect. 4 shall I come to some recent results.


\newsec{A short review of the main protagonists}
\subsec{Conformal Field Theories}
\nind
Rational conformal field theories (cft) are constructed out of
representations of the product $\CA\otimes \CA$  
of two copies of an algebra $\CA$ \ \refs{\BPZ{--}\MoSe}. The algebra
  $\CA$ is the maximally extended ``chiral" algebra, 
generated by the product of holomorphic fields in the theory. 
It contains the Virasoro algebra as a subalgebra 
(or a subalgebra of its enveloping algebra)
; examples are provided 
by the Virasoro algebra itself or its supersymmetric extensions,  by 
affine (or ``current") algebras $\widehat{\gg}$,  by the so-called 
$W$ algebras, \dots
A central role is played by the theories with an affine 
algebra $\widehat{\gg}$, as it seems likely that all the others may be 
obtained from them by the so-called coset construction \GKO. 
The data that specify such a cft (for a given  $\CA$) are 
\eqn\Ia{ \{k\,, \  (\Gl_i,\Gl_{\bi})\, , \ C_{IJK}\}}
where $k$  
is a generic notation for
the central charge(s) of $\CA$; 
$(\Gl_i,\Gl_{\bi})$ describe the set of irreducible representations
of $\CA\otimes \CA$ that appear in the decomposition of 
the Hilbert space of the theory 
\eqn\Ib{ \CH= \oplus_{I=(i,\bi)} V_{\Gl_i,k}\otimes V_{\Gl_{\bi},k}\ }
where $V_{\Gl}$ is an irreducible representation of $\CA$ of highest weight 
(h.w.) 
$\Gl$. Here and in the following, a capital like $I$ denotes a pair
of indices $(i,\bi)$. The vacuum state plays a special role in the
theory: it appears in a representation denoted $\Gl=\bar\Gl=\un$.

If $\CA\ne$~Vir such a $V_{\Gl}$ 
splits into a finite or infinite sum of 
irreducible h.w. representations of the Virasoro algebra~:
$ V_{\Gl}=\oplus V_{h_i}^{{\rm (Vir)}}$, and the $h_i$, eigenvalues of the 
dilatation generator $L_0$, are called conformal weights. 
 Finally, in \Ia, $ C_{IJK}$ are the structure constants of the
operator product algebra (OPA) of the  ``primary'' fields 
(highest weight representations of $\CA\otimes\CA$)
\eqn\Ic{\Phi_I(z,\bar z) \Phi_J(w,\bar w)= \sum_K C_{IJK} 
(z-w)^{h_k-h_i-h_j}(\bar z-\bar w)^{h_{\bar k}-h_{\bi}-h_{\bj}}
\{ \Phi_K(w,\bar w) +\CO(|z-w|) \} }
with $h_i, h_{\bi}$ the conformal weights of $\Phi_I$. In the simplest
case of $\CA= {\rm Vir}$, these data reduce to
$\{c,(h_i,h_{\bi}),C_{IJK}\}$, $c$ the Virasoro central charge. 

These data are strongly constrained by consistency conditions in the
matching between the  ``left'' (holomorphic) and ``right'' 
(antiholomorphic) components of the theory. This is exploited in two 
parallel procedures that deal with the 0-point function in genus
1 and with the 4-point function in genus 0 and that 
determine respectively the set 
$ (\Gl_i,\Gl_{\bi})$ and the $ C_{IJK}$.

\medskip\nind
{\sl Modular invariance of the partition function on a torus}
\penalty 20000\par \penalty 20000 
\noindent
One considers first the genus 1 0-point function of the cft on the
torus $\IC / \IZ \omega_1 \oplus \IZ \omega_2$. It may be written as
\eqn\Id{Z(\tau)=\tr_{\CH}\, q^{L_0-c/24}\, {\bar q}^{\bar L_0-c/24} }
where $L_0,\bar L_0$ are the Virasoro (dilatation) generators, 
$\tau=\omega_2/\omega_1$ is the modular ratio of the torus 
and $q= \exp 2i\pi \tau$ is its nome. According to \Ib, this 
may be rewritten as \JLC
\eqn\Ie{Z= \sum_{(\Gl_i,\Gl_{\bi})} N_{\Gl_i,\Gl_{\bi}}
\chi_{{{}\atop\Gl_i}}\!(q) \chi_{{{}\atop\Gl_{\bi}}}\!(\bar q) .}
$N_{\Gl_i,\Gl_{\bi}}$ is the multiplicity of occurrence of
$\Gl_i,\Gl_{\bi}$ in \Ib, hence a non negative integer and 
\eqn\Iee{ \chi_{{{}\atop\Gl}}(q)= \tr_{V_{\Gl,k}} q^{L_0-c/24} }
is the character of the $(\Gl,k)$ representation of the $\CA$ algebra, 
a generating function of the number of independent 
states of that representation graded by the eigenvalues of $L_0-c/24$. 
The unicity of the vacuum is expressed by 
\eqn\Ieea{N_{\un\,\un}=1 \ .}
The 
partition function $Z$ is required to be intrinsically attached to the torus, 
thus be independent of $SL(2,\IZ)$ redefinitions of its periods   
\refs{\JLC{--}\WN}. 
In other words, it must be a modular invariant function of $\tau$
\eqn\If{ Z(\tau)=Z\({a\tau+b\over c\tau +d}\), \qquad a,b,c,d \in \IZ, \ 
ad-bc=1\ .}
This leads to the following \par
\smallskip\nind
{\bf Problem 1~:} {\sl For a given algebra $\CA$, classify all sesquilinear 
forms in the characters \Ie\ with  non negative integer coefficients
and $N_{\un\,\un}=1$.  } 
\par\nind
Solution to this problem determines the representation content \Ib.
\medskip
%
\medskip\nind
{\sl Monodromy invariance of the 4-point function}\nind\penalty 10000
After use of a M\"obius transformation mapping three of the
arguments to 0, 1 and $\infty$, a correlation function of 4 primary fields 
may be expressed as  \BPZ
\eqna\Ig
$$
\eqalignno{\bra \Phi_I(\infty) \Phi_J(1)\Phi_K(z,\bar z) \Phi_L(0)\ket 
&= \sum_{M=(m,\bar m)} C_{IJM}C_{KLM} \CI_m(z) \CI_{\bar m}(\bar z)\qquad
&\Ig a \cr
&= \sum_{N=(n,\bar n)} C_{IKN}C_{JLN} \CK_n(1-z) \CK_{\bar n}(1-\bar z)
\ .\qquad &\Ig b \cr }$$
\vskip-10mm
$$  {\rm written\ symbolically\ as\ }\qquad 
{\epsfxsize=3cm \epsfbox{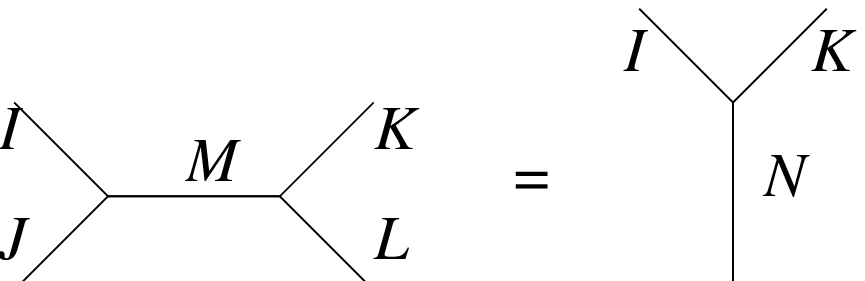}} $$
%
\smallskip
\noindent The so-called conformal blocks $\CI_m(z)$ behave as $\CI_m(z)\approx 
z^{h_m-h_k-h_l}$ as $z\to 0$ and likewise for $\CK(1-z)$ close to $z=1$.
They have non trivial
monodromy as $z$ circles around 0, 1 or $\infty$. Locality requires 
that the function be single-valued, hence that the combination \Ig{a}\
be monodromy invariant, or equivalently, that \Ig{a} and \Ig{b} be consistent.
Hence the \par
\smallskip\nind
{\bf Problem 2~:} {\sl Find all the monodromy invariant combinations of 
conformal blocks \Ig{a}.}\par\nind
This determines the set of structure constants $C_{IJK}$. 


\subsec{Integrable height lattice models}
\nind Height or {\it solid-on-solid} models are statistical mechanics
models originally designed to describe a fluctuating surface or 
interface with no overhang~: to each site $i$ of a two-dimensional
lattice is attached
an integer $a_i$, describing the height of the surface above a reference 
plane. In the so-called restricted solid-on-solid models (RSOS), additional
constraints are put on the range of $a$~: $1\le a\le n$, and on its
fluctuations from  
site to site, typically $|a_i-a_j|=1$ if $i$ and $j$ are neighbours. 
Supplemented with an appropriate Boltzmann weight, this is the integrable
RSOS model of \ABF. It is integrable in the sense that its 
Botzmann weights satisfy
the Yang-Baxter equation and it possesses several 
critical regimes, in which its correlation length diverges. Since we are 
mainly interested in the connection with conformal theories, we shall
concentrate on one of these critical domains. The constraints above have been
reinterpreted in \VPun\ as follows~: heights are regarded as the vertices 
of the graph 
$\buildrel 1 \over \bullet \!\!{{}\over{\qquad}}\!\!
\buildrel 2 \over \bullet 
\!\!\cdots\cdots \buildrel n \over \bullet $, 
the second constraint expressing that neighbouring sites on the lattice
are mapped onto neighbouring vertices on the graph. It is quite natural to
try to generalize this picture to more general graphs. 
The search of critical and
integrable models within this framework has to be supplemented by a certain
Ansatz on the form of the Boltzmann weights,  
built out of representations of some algebra $\CT$, typically some quotient 
of the Hecke algebra (a deformation of the algebra of the symmetric
group) acting on the space of paths on the graph. The simplest case
is given by the Temperley-Lieb algebra \Bax. 
I shall not be more explicit on this
point, referring the reader to the literature {\TL}. 
So within such an 
Ansatz, we have to face the
\smallskip\nind
{\bf Problem 3 : }{\sl Find all the graphs that support a representation
of the algebra $\CT$ and hence lead to a critical integrable RSOS model. }


\subsec{Topological Field Theories}
\noindent In topological --better called cohomological-- field theories (tft) 
one is dealing with 
a finite set of fields $\phi_i$, $i=1,\cdots,n$ living 
(in a two-dimensional case) on some Riemann surface $\Sigma$ of genus $g$.
These fields are in the cohomology space of a certain nilpotent 
operator $Q$, $Q \phi =0$. One is interested in the correlation 
functions
\eqn\Ih{\bra \phi_{i_1}\cdots  \phi_{i_k}\ket_{\Sigma} \ . }
One assumes that 
all correlators between the $\phi$'s and a $Q$-exact field vanish. An 
example of such a $Q$-exact field is provided by the energy-momentum
tensor that describes as usual the response of the theory to a 
change of metric~: $T=\{Q, G\}$.  As a result of these axioms, 
the correlator \Ih\ is insensitive to the representative of $\phi_i$
in its cohomology class and insensitive to a change of metric~: it is 
a topological invariant that depends only on the labels $i_1,\cdots,i_k$ 
and on the genus $g$ of $\Sigma$. In fact, one is considering {\it families}
of such correlation functions, depending on $n$ moduli 
$t_1, \cdots, t_n$
in one-to-one correspondence with the fields $\phi_1,\cdots,\phi_n$.   
(What is the actual topological meaning of 
this object remains to be seen and is a matter of a case by case analysis.)

One then demands  that these correlation functions 
satisfy a certain number of constraints:
\nind
(i)  the 3-point genus 0 function $\bra \phi_i\phi_j\phi_k\ket_0 
=C_{ijk}(\tp)$
satisfies the integrability conditions that enable one to write it 
as  
\eqn\Ii{   C_{ijk}(\tp)={\partial^3 F\over \partial t_i
\partial t_j \partial t_k }\qquad 
{\rm for\ some\ function\ }F(\tp) \ ;} 
\nind
(ii) the 2-point genus 0 function $\bra \phi_i\phi_j\ket_0=\eta_{ij}=C_{1ij}$ 
is an invertible, $t$-independent metric and may be used to raise and lower 
the indices of $C_{ijk}$;\nind
(iii) the 3-point functions satisfy the condition of factorization 
\vskip-10mm
\eqn\Ij{ {\epsfxsize=35mm \epsfbox{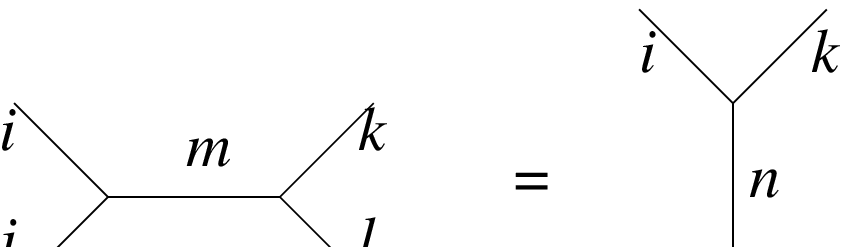}}\qquad{i.e.}\qquad
\sum_m  C_{ijm} C^m_{\ kl} =\sum_n  C_{ikn} C^n_{\ jl}\ .}
(iv) One usually adds the condition 
that $F(\td)$ is a quasihomogeneous function of the $t$'s. 
These constraints result in a set of non-linear partial differential
equations for $F$, the so-called Witten-Dijkgraaf-Verlinde-Verlinde (WDVV)
equations \refs{\Wi{--}\DVV}.
\smallskip\nind
{\bf Problem 4 :} {\sl Find solutions to these WDVV equations.} \nind
\smallskip
As such, the problem is  quite vast and it is remarkable that 
results may be obtained in such a general framework \Dub.  
For our purpose, however, it 
is still too general and has to be supplemented 
with a certain Ansatz on the tft. In the following this will be provided 
by the connection with $N=2$ theories obtained by the Kazama-Suzuki
coset construction. 

$N=2$ superconformal theories are known to provide a class of tft's, after
a modification called twisting. They are conformal theories endowed
with a larger algebra, generated by the energy momentum tensor $T(z)$,
two supersymmetry generators denoted $G^{\pm}(z)$ and a $U(1)$ 
current $J(z)$. I shall not write the algebra here, referring the reader 
to the literature \Wa. Suffice it to say that the moment
$G^+_{-\oh}=\oint {dz\over 2i\pi}G^+(z)$ is a nilpotent operator that 
we shall denote $Q$ for brevity. The fields in the cohomology space of $Q$
are called {\it chiral}\foot{an unfortunate terminology, 
not to be confused with the concept of chiral algebra.}
; their eigenvalues for $L_0$ and $J_0$ satisfy $h=\oh q$.
The twisted energy-momentum $T^{{\rm top}}(z)=T(z)+\oh J(z)$
is $Q$-exact: $T^{{\rm top}}(z)=\oh \{Q,G^-(z)\}$ and the correlators
of the chiral fields $\phi$ satisfy the axioms above. 
Deformations are introduced by perturbing the action by
\eqn\Ik{ \Delta S=\sum_{l=1}^n t_l \int \d^2w\, G^-_{-\oh}{\bar G}^-_{-\oh} 
\phi_l(w,\bar w)+ {\rm c.c.} }
and are shown to preserve the topological character of the correlators. 
Thus this procedure constructs a large class of 
solutions to the WDVV equations.

In turn, $N=2$ theories may be constructed by a supersymmetric
version of the coset construction of \GKO, due to Kazama and 
Suzuki. 
These theories are based on a pair
$(G,H)$ such that $G$ is a simple compact Lie group, $H$ a
subgroup with rank~$G$ = rank~$H\ +2n$ and $G/\(H\times
U(1)^{2n}\)$ is a k\"ahlerian manifold \KS. 
(In the ordinary coset construction, 
 they are described by the coset
$(G\times SO(\dim G/ H))/H$.) Those that will
concern us are the grassmannians $SU(n+m)_k/\(SU(n)_{k+m}\times SU(m)_{k+n}
\times U(1)\)$, where the label denotes the central charge 
of the affine algebra. Their Virasoro central charge is $c=3nmk/(n+m+k)$.



\newsec{The case of $SU(2)$}

\subsec{Classification of modular invariants}
\nind
The classification of modular invariant partition functions has 
been completed for theories with a $\widehat{ su}(2)$ affine algebra
as a chiral algebra, or for some of their relatives through the
coset construction. It is convenient to label the representations 
of $\widehat{ su}(2)$ by the integer $\Gl=2j+1$, if $j$ denotes the spin
of the ``horizontal'' algebra. If the central charge is $k$, 
then $1 \le \Gl\le k+1$. One finds after a fairly strenuous 
analysis that all modular invariant $Z$ are in one-to-one  correspondence 
with {\it simply laced } simple Lie algebras (of $ADE$ type) \CIZ. Indeed
if one singles out the diagonal terms in  $Z$, writing
\eqn\IIa{Z=\sum_{\Gl\in\{m_i\}}  |\chi_{\Gl}|^2 +\hbox{non diagonal terms}}
then one finds that the diagonal labels $\Gl$ run over the {\it exponents}
$m_i$ of an $ADE$ algebra of Coxeter number $h$ related to the level $k$ by
$h=k+2$. (Recall that the Coxeter number $h$ and exponents $m_j$ 
are integers $ 1\le m_j \le h-1$ such that the
eigenvalues of the Cartan matrix $C$ of an
$ADE$ algebra are $4\sin^2{\pi m_i\over 2h}$ or equivalently, those
of the adjacency matrix of the Dynkin diagram $G=2 
\II-C$ are 
$2 \cos {\pi m_i\over h}$.)


\subsec{Determination of the structure constants}
\nind This program has also been carried through for $\widehat{ su}(2)$ 
models and the corresponding ``minimal'' conformal theories \DFZa. The 
$C_{IJK}$ are complicated, generically transcendant numbers, but somehow
they also know about the $ADE$ pattern~! This is apparent if, for
two $\widehat{ su}(2)$ theories with the same $h=k+2$, one forms
the {\it ratios} of structure constants    pertaining to spin 
zero (left-right symmetric) fields. One finds that \PZ
\eqn\IIb{{C_{(ii)(jj)(\ell\ell)}^{(D)\ {\rm or\ } (E)}\over 
C_{(ii)(jj)(\ell\ell)}^{(A)} }= \sum_a 
{\psi^{(i)}_a\psi^{(j)}_a\psi^{(\ell)\,*}_a\over\psi^{(1)}_a}\ .}
Here $\psi^{(i)}_a$ is the $a$-th component of the $i$-th 
orthonormalized eigenvector of the adjacency matrix of the $D$ or $E$ 
Dynkin diagram. 
The numbers that appear on the right hand side are algebraic 
(in fact they are square roots of rationals!) and  form the structure
constants of an associative and commutative algebra, first introduced
in \VP\ in the context of lattice models. (For the $A$ Dynkin diagram, this 
reduces to the Verlinde fusion algebra.) 

Formula \IIb\ does not seem to be a trivial consequence of the 
crossing equations \Ig{}. One has to go through the painstaking 
determination of all the $C$'s to check it. On the other hand, as
anticipated in \VP, \IIb\ is easily derived within the lattice models
to be discussed now \PZ.


\subsec{Height models on a graph}
\nind
Although it may not be manifest, the RSOS models that  have  been
introduced above have to do with the $SU(2)$ group. This may be
seen through the chain of connections \nind 
\quad $XXZ$ ($SU(2)$) spin chain $\leftrightarrow$ 6 vertex model 
$\rightarrow$   RSOS model of ABF $\rightarrow$  new models
\nind
or at a more technical level, on the fact that the Temperley-Lieb
algebra on which the construction relies is related to the 
$\CU_q\,sl(2)$ quantum group \TL. In the continuum limit, the RSOS models
 are
described by these minimal coset theories $SU(2)_k\times SU(2)_1/SU(2)_{k+1}$.

In his analysis of this class of height models attached to graphs, 
Pasquier showed that the condition of criticality is that the largest 
eigenvalue $\gamma_{{\rm max}}$
of the adjacency matrix be less or equal to 2. Leaving 
aside the case where it equals 2, which is a marginal case (related to
conformal theories with $c=1$), we recall the well known fact that 
the only  graphs with $\gamma_{{\rm max}}<2$
are the $ADE$ Dynkin diagrams (plus the $\IZ_2$ orbifold
of the $A_{2n}$ graph, that does not produce any new lattice model). 

Hence in this context, we see that the $ADE$ diagrams have been 
singled out by a spectral condition on their adjacency matrix.


\subsec{Topological Field Theories}
\nind 
The simplest $N=2$ ``minimal" theories, constructed through the coset
$\(SU(2)\times  SO(2)\)/U(1)\cong
\(SU(2)/U(1)\)\times U(1)$, and the ensuing topological field theories,
may be classified following the modular invariant procedure. One finds 
again an $ADE$ classification, with some decoration associated to the 
$U(1)$ factors \Gep. These decorations, however, do not affect the
chiral content of these theories, and one finds \TN\ that all the 
chiral primary fields are spinless and have a $U(1)$ charge 
given by ${m_i\over h}$, where $h$ and $m_i$ are as before the 
Coxeter number and exponents.

Another way of seing this $ADE$ classification at work is through the
Landau-Ginsburg approach. It has been shown that a class of $N=2$
theories admits a description in terms of a superpotential. The latter
is a function of some chiral superfields $X,Y,\cdots$ only, and has been 
argued to be  a quasi-homogeneous polynomial $W(X,Y,\cdots)$ 
in these fields, 
with an isolated critical point at the origin: $W'_X|_0=W'_Y|_0=0$. 
It is thus to be found in  lists of singularities \MVW.
In particular, for the minimal theories, the superpotential must 
have no modulus and is among the well known $ADE$ singularities \AGZV
\eqnn\IIc
$$\eqalignno{A_n \qquad W&= {X^{n+1}\over n+1}\cr
	  D_{n+2}\qquad W&= {X^{n+1}\over 2(n+1)}+ XY^2\cr
	     E_6 \qquad W&= {X^3\over 3}+ {Y^4\over 4}& \IIc \cr
	     E_7 \qquad W&= {X^3\over 3}+ {XY^3\over 3}\cr
	     E_8 \qquad W&= {X^3\over 3}+ {Y^5\over 5}\ .\cr }$$
The $U(1)$ charges of the chiral fields, hence the gradings of the 
fields or their conjugate $t$ parameters in the tft, are then read 
off the degrees of the local ring of the singularity; as is well
known, they reproduce the Coxeter exponents of the $ADE$ algebras \AGZV.

We have thus found two independent routes to the ADE classification. 
The latter sheds some light on the former. Unfortunately it does not 
generalize easily to higher rank cases, due to the fact that many
$N=2$ theories escape the Landau-Ginsburg description. 

This makes a third approach quite valuable. A year and a half ago, Dubrovin
observed that one can associate a solution to the WDVV equations with each
finite Coxeter group. Recall that Coxeter groups are generated by
reflections in $(n-1)$-dimensional hyperplanes in a $n$-dimensional
 Euclidean space $V$; the hyperplanes are orthogonal (through the origin)
to a set of $n$ independent vectors $\Ga_a$ called roots. 
Finite Coxeter groups are classified: beside the Weyl groups of the simple
Lie algebras, $A_p$, $B_p$, $C_p$, (the two latter Coxeter groups 
being isomorphic), $D_p$, $E_6$, $E_7$, $E_8$, $F_4$ and 
$G_2$, there are the groups $H_3$  and $H_4$ of symmetries  
of the regular icosaedron and of a regular 4-dimensional polytope, and 
the infinite series $I_2(k)$ of the reflection groups 
of the regular $k$-gones in the plane. In general, if we use the roots 
$\Ga_a$ as a basis, in the reflection $S_a$ in the hyperplane orthogonal to 
$\Ga_a$, a vector $x=\sum_c x_c \Ga_c$  is transformed into 
\eqn\IId{
\eqalign{S_a\ :\ x\mapsto 
x'&=\sum x'_c \Ga_c=x-{2(\Ga_a,x)\over(\Ga_a,\Ga_a)}\Ga_a \cr
 & \cases{ x'_a=-x_a+2 \sum_{c\ne a} G_{ac} x_c & \cr
  x'_b=x_b & if $b\ne a$\cr} \cr}}
with $G=2 \II -C$ as in sect. 3.1.
In Dubrovin's work, the homogeneity degrees 
of the variables $t_i$ and of $F$ are respectively $1-(d_i-2)/h$ and 
$2+2/h$ where $h$ is the Coxeter number of the group $G$
and $d_i$ are the degrees of the $G$ invariant polynomials in the coordinates
of $V$. 
The $ADE$ solutions are the previous minimal solutions. What are the other 
solutions, associated with non $ADE$ finite Coxeter groups?
They cannot be obtained by twisting of a consistent, modular invariant 
$N=2$ theory. It seems that when coupled to gravity, only the $ADE$ theories
are consistent at higher genus \EYY.

Does the same pattern with all the Coxeter groups appear in the
conformal context? The answer is yes, although the reason remains
elusive. 
By inspection, one may check that the Coxeter groups label
a certain class of subalgebras of the OPA of the $ADE$ theories \Zub. 
This class is defined as obeying two  constraints: i)
the subalgebra is generated by spinless (left-right symmetric)
fields; ii) it contains  the identity and the field of largest 
$\Gl$, i.e.   $\Gl=h-1$, $h$ the Coxeter number of the $ADE$ algebra. 
Whether the latter constraint is merely technical or reflects
something deeper remains to be 
seen.  Note that again among all the theories endowed with these 
operator algebras, only the $ADE$ ones are consistent (modular invariant)
at non zero genus.



\newsec{The case of $SU(N)$}
\subsec{Classification of $\widehat{su}(3)$ modular invariants}
\nind Let $\GL_1,\cdots,\GL_{N-1}$ be the fundamental weights 
of $su(N)$ and $\rho=\GL_1+\cdots +\GL_{N-1}$ be their sum.
Let $\CP_{++}^{(k+N)}$ denote the set of integrable weights 
(shifted by $\rho$) of the affine algebra $\widehat {su}(N)_k$ 
at level $k$, (the ``Weyl alcove"),
\eqn\IIIa{\CP_{++}^{(k+N)}=\{\Gl=\Gl_1\GL_1+\cdots 
+\Gl_{N-1}\GL_{N-1}| \Gl_i\ge 1,\ \Gl_1+\cdots \Gl_{N-1}\le k+N-1\} \ .}
exemplified on $\widehat{su}(3)$ at level 3 in Fig. 1.
\fig{$\CA^{(6)}=\CP^{(6)}_{++}$}{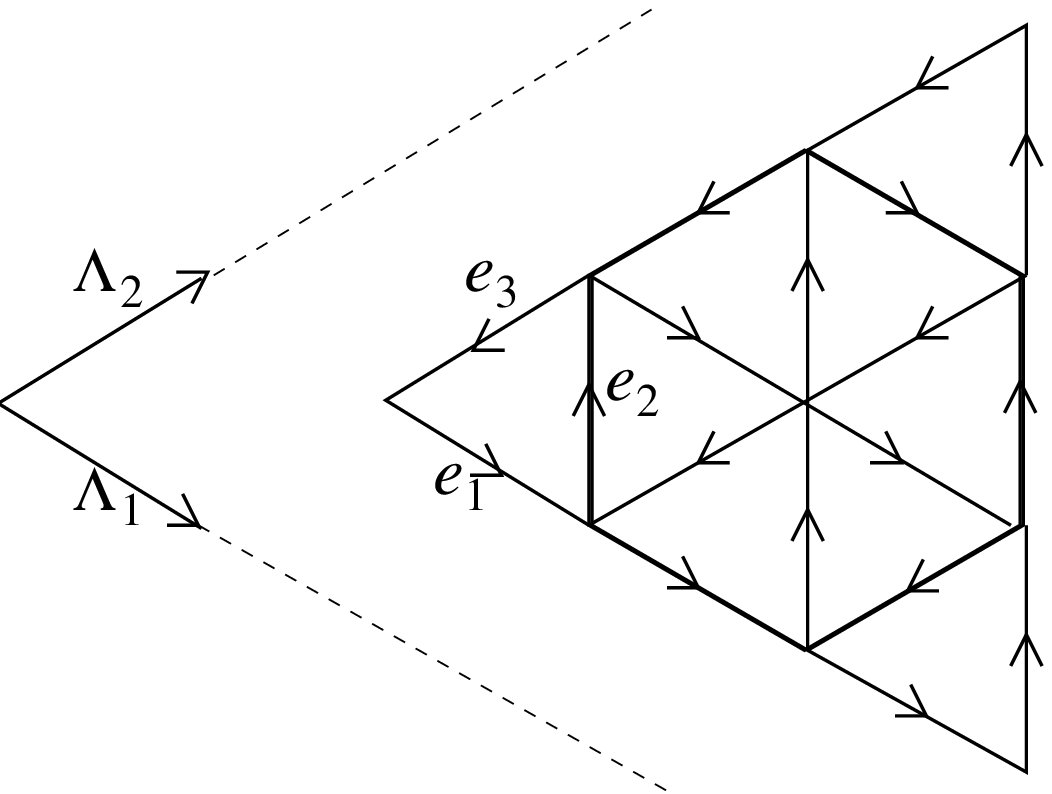}{3cm}\figlabel\triang

The partition function $Z$ of the $\widehat{su}(N)_k$ theory 
must be a sesquilinear form in the characters labelled by some integrable
weights 
\eqn\IIIaa{Z=\sum_{\Gl}  |\chi_{\Gl}|^2 +\hbox{non diagonal terms}}
and the representation $\Gl=\bar\Gl=\rho$ 
must appear once and only once (see \Ieea). In the case of $\widehat{su}(3)$, 
this is a problem with a fairly long history \BALZ, 
that culminates with the recent proof by Gannon that the list of known 
invariants is complete \Gan. The classification
 features again a diagonal series, several
infinite series obtained by orbifolding or twisting of the former, and a 
few exceptional cases. This leaves us with the question~: can one
find a (geometrical or otherwise) meaning to this list~? By extension
of the $su(2)$ case, we shall call {\it exponents} the weights $\Gl$
that label the diagonal terms in $Z$ and denote their set by $\Exp\,$.

%
\subsec{Construction of $SU(3)$ lattice models }
\nind
The basic model -- the analogue of the Andrews-Baxter-Forrester model --
has heights $\Gl$ living in $\CP_{++}^{(k+N)}$, the set of integrable weights 
at a given level, and Boltzmann weights given by a representation of
(a quotient of) the Hecke algebra \refs{\JMO{--}\We}.
This may be regarded as a model on the graph 
$ \CA^{(k+3)}=\CP_{++}^{(k+N)}$ whose  bonds are oriented along the
three vectors of vanishing sum $e_1=\GL_1$, $e_2=\GL_2-\GL_1$ and
$e_3=-\GL_2$ (see fig. 1). Its vertices may be 3-coloured, i.e. assigned a
$\IZ_3$ grading $\tau(.)$ 
(the ``triality'') in such a way that for  the adjacency matrix $G$, 
$G_{ab}\ne 0$ only if $\tau(b)=\tau(a)+1\  {\rm mod\, }3$. This 
adjacency matrix has eigenvalues 
\eqn\IIIb{\Gc^{(\Gl)}=\sum_{i=1}^3 \exp {2i\pi\over h} (e_i,\Gl)\ ,}
with $\Gl$ running over the set $\CP_{++}^{(k+N)}$, 
$(.,.)$ is the Cartan inner product in weight space and $h= k+3$.  
It is easy to see that the transposed matrix $G^t$ that describes 
the graph with all orientations reversed commutes with $G$. Thus $G$ 
is a normal matrix, that may be diagonalized in an orthonormal basis. Its
diagonalization is indeed provided by the Verlinde formula, since $G$ and 
$G^t$ are the matrices of fusion by the two fundamental 3-dimensional 
representations of $SU(3)$.

Guided by the analogy with the case of $SU(2)$, one may look for
graphs $\CG$ satisfying the following properties:
\item{i)} $\CG$ is 3-colourable, i.e. 
a $\IZ_3$ grading $\tau$ may be assigned to
its vertices, such that $G_{ab}\ne 0$ only if $\tau(b)=\tau(a)+1\ \mod 3$;
\item{ii)} the graph is invariant under an involution $a\mapsto \bar a$
such that $\tau(\bar a)=-\tau(a)$ and $G_{\bar a \bar b}=G_{ba}$; 
\item{iii)} its adjacency matrix is normal: $[G,G^t]=0$;
\item{iv)} the spectrum of eigenvalues of $G$ is a subset $\Exp$ 
(with possible multiplicities)
of the set \IIIb, for some $h$. In other words, this
spectrum is described by a  set of ``exponents'' $\Gl \in \CP^{(h)}_{++}$
 occuring with a multiplicity $N_{\Gl}$~;
\item{v)} these exponents match the exponents of one of the $SU(3)$ coset
modular invariants; in particular,  $\rho$ belongs to the exponents, 
with multiplicity one. \nind

$$   {\epsfxsize=6cm \epsfbox{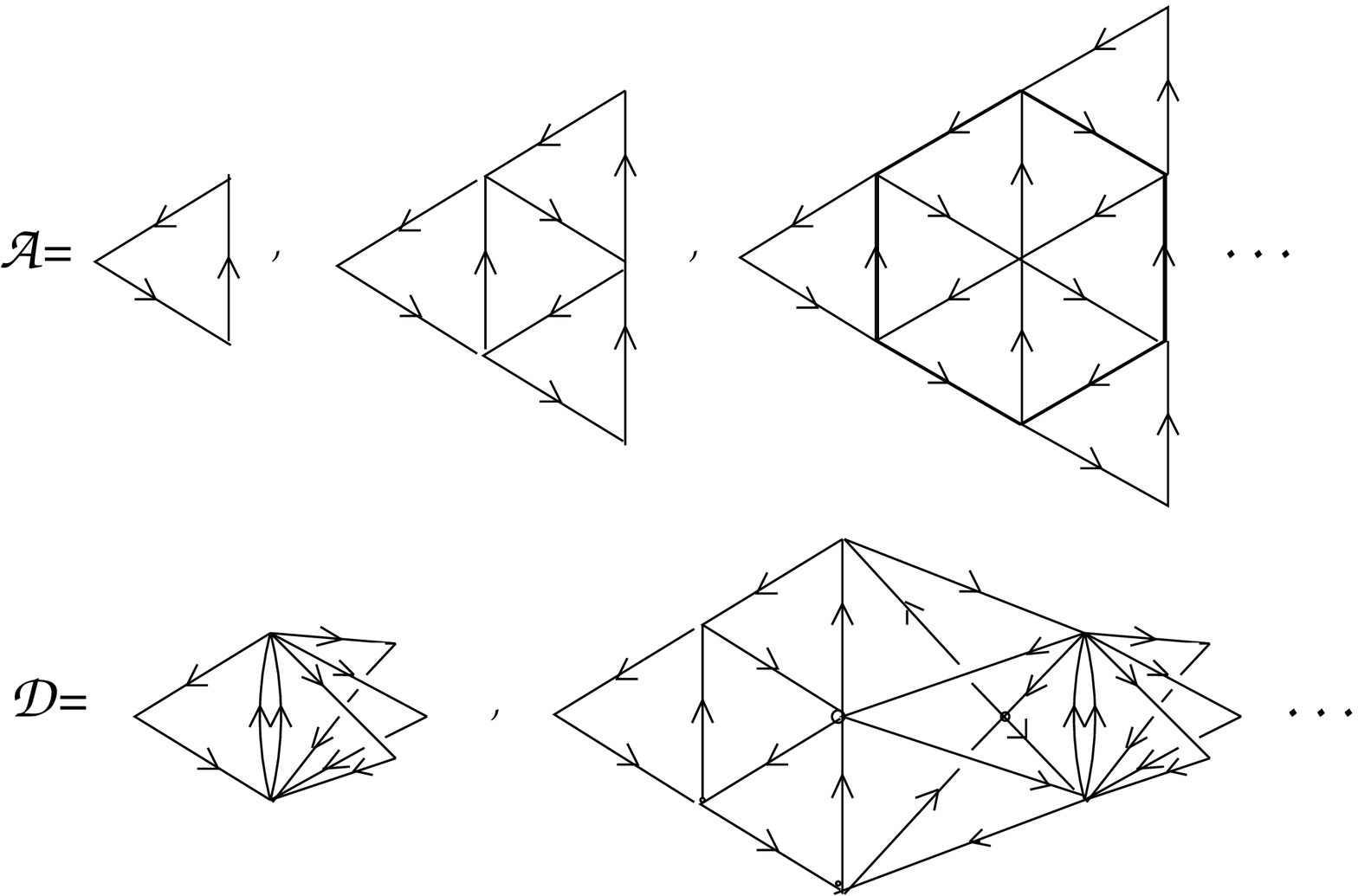}} $$
\vskip-5mm
\fig{Some of the $SU(3)$ graphs. (Not all orientations of edges have been
shown on the last one.)}{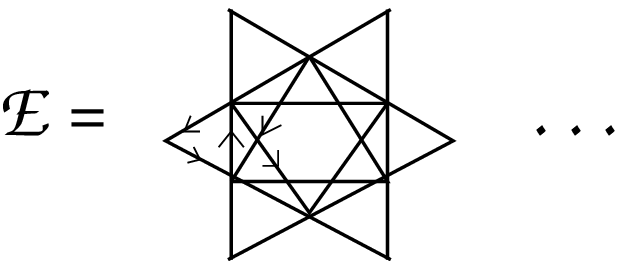}{30mm}\figlabel\triang

In \Ko\ and  \DFZ, other solutions  than the graphs $\CA$ above were found; 
in fact we
found enough solutions to match all the modular invariants 
(of coset theories) (see Fig.\triang). They come again in infinite series
called $\CA$ and $\CD$ and a few exceptional cases $\CE$; 
these notations are suggested by the analogy with the pattern of the 
previous case of $SU(2)$, but it should be stressed that there is no
relation with the ordinary $ADE$ scheme. We 
checked in almost all cases that these graphs support a representation of the 
appropriate Hecke algebra, hence an integrable $SU(3)$ height model. 
(Contrary to the case of $SU(2)$, this construction had to be done case by 
case and was more and more cumbersome as the complexity of the 
graph increased \refs{\DFZ{--}\So}.) 
It is believed that the continuum limit of 
these lattice models reproduces the corresponding (coset) cft's. 
In \DFZ, the construction of the graph corresponding to a given 
modular invariant was carried out in an empirical
way, essentially by trials and errors%
. We now have some indications that there exists
a direct procedure for determining the graph \lref\PZn{V. Petkova and J.-B. 
Zuber, in preparation.} \PZn.

\subsec{TFT's based on $SU(3)$}
\nind  By the Kazama-Suzuki construction, one knows that the cosets
$$ {SU(3)_k\over SU(2)_{k+1}\times U(1)}\times SO(4)_1 $$
lead to $N=2$ theories  and, after twisting, to  topological
field theories.
As before for the minimal case, the classification of $SU(3)$ modular
invariants reflects itself on that of these coset theories and the
resulting tft's. A question is: what is the distinctive feature 
of tft's emanating from this series of cosets? 

\newsec{New Coxeter groups}
\nind
With each graph $\CG$ of the previous type, of adjacency matrix $G_{ab}$, 
we can associate a Coxeter group as follows. We consider a vector space 
$V$ on $\IR$, with a basis $\{\Ga_a\}$. We define  a bilinear 
symmetric form on $V$ by 
\eqn\IIIm{\eqalign{(\Ga_a,\Ga_a)&=1 \cr
		(\Ga_a,\Ga_b)&=\oh \(G_{ab}+G_{ba}\)
\qquad {\rm if}\  	\ a\ne b\ . \cr	}}
This allows to define the reflexion $S_a$ 
\eqn\IIIn{S_a\quad : \quad x\mapsto x'=x-2(\Ga_a,x) \Ga_a}
or in terms of the components $x_b$~: $x=\sum_b x_b \Ga_b$
\eqn\IIIo{\eqalign{x'_a&=-x_a-\sum_c(G_{ac}+G_{ca}) x_c \cr
x'_b&= x_b  \qquad {\rm if}\  	\ b\ne a\ . \cr	}}
Note the change of sign with respect to the case considered above 
\IId. 

Let us denote $\tilde G_{ab}=(G_{ab}+G_{ba})$.
The following properties are standard or easily proved:
\item{(i)} $S_a^2=\un\ .$
\item{(ii)} If $\tilde G_{ab}=0$, then $S_a$ and $S_b$ commute and 
the product $S_a S_b$ is of order 2
$$ (S_a S_b)^2=\un \ .$$
\item{(iii)} If $\tilde G_{ab}=1$, then $S_a S_b$ is of order 3 and 
more generally, if $\tilde G_{ab}=
2\cos {\pi p\over q}$, with $p$ and $q$ coprime integers, 
$S_a S_b$ is of order $q$. 
\item{(iv)} With $h$ defined as in \IIIb,  
the bilinear form $(\Ga_a,\Ga_b)=\Gd_{ab}+\oh\(G_{ab}+G_{ba}\)$
is positive definite for $h<6$, semi--definite for $h=6$ and 
indefinite for $h>6$.

These properties are readily established. Less trivial is the following 
proposition, that extends to the present case the well-known 
relationship between the exponents of finite Coxeter groups and the
eigenvalues of the ``Coxeter element''. 
Let 
\eqn\IIIp{R = \prod_{\tau(a)=0} S_a  \prod_{\tau(b)=1} S_b  
\prod_{\tau(c)=2} S_c \ , }
obtained by the product of the three blocks of reflections 
of given triality. Then

\nind {\bf Proposition : }{\sl The element $R$ is independent, up to
conjugation,  of the order
of the blocks and of the order  of the $S$ within each block,  
and its spectrum is of the form
\eqn\IIIpro{
 -\exp 3 {2i\pi\over h}(e_i,\Gl)\qquad \Gl\in \Exp\, \CG, \quad i\ 
{\rm fixed: }\  1\le i \le 3 }
with the same notations as in \IIIb. In particular this set of eigenvalues
is independent of $i=1,2,3$, up to a permutation. }

The proof relies on a simple extension of the original proof 
by Coxeter of the analogous statement for finite Coxeter groups. 
By a possible reordering of the vertices of the graph assume that
the matrix $G$ takes the block form
\eqn\IIIq{G=\pmatrix{0 & A & 0 \cr 0 & 0 & B\cr C & 0 & 0 \cr}\ .}
Then the numbers $\Ge_i^{(\Gl)}:= \exp{2i\pi\over h}(e_i,\Gl)$ are
the roots of the polynomial 
\eqn\IIIpa{
\eqalign{ \prod_{\Gl} (z-\Ge^{(\Gl)}_1)&(z-\Ge^{(\Gl)}_2)(z-\Ge^{(\Gl)}_3) 
	= \det(z^3-z^2 G+z G^t-\un)\cr
	&= \det\pmatrix{(z^3-1)\un & -z^2 A & z C^t \cr
			 z A^t & (z^3-1) \un & -z^2 B \cr
			 -z^2 C & z B^t & (z^3-1)\un \cr} = \Delta (z^3) }}
which is in fact a polynomial in $z^3$ as is readily seen by multiplying 
or dividing the rows and columns by  the appropriate power of $z$. In fact
\eqn\IIIpb{\Delta(z^3)=\det (z^3 T -T^t)}
where 
\eqn\IIIpc{T=\pmatrix{\un & -A & C^t \cr
		       0 & \un & -B \cr
		       0 & 0 & \un\cr }	\ .}
Then it is easy to prove that the  ``Coxeter element'' of
\IIIp\ is conjugate to $-T^{-1}T^t$ and has thus by the 
previous discussion the spectrum $\{-\Ge^{(\Gl)\,3}\}$. 
Finally, the independence up to conjugacy with respect to the order
in \IIIp\ follows from the fact that each block in \IIIp\ has square 
one.

According to the point (iv) above, the groups generated by the $S_a$
is finite for $h<6$, and must therefore identify with one of the 
well known finite Coxeter groups. One verifies indeed that the 
groups associated with the graphs $\CA^{(4)}$ 
(the $SU(3)$ weights of level 1), 
$\CA^{(5)}$ (the same at level 2) and $\CA^{(4)}$ in 
which all the links carry $G_{ab}=2\cos{\pi\over 5}$ 
coincide respectively with the finite groups $A_3$, $D_6$ and
$H_3$ of orders 24,  $2^5 6!$ and 120. The first two identifications 
could have been anticipated from identifications between coset realizations 
of $N=2$ theories. Indeed (see for instance \DFLZ)
\eqn\IIIr{\eqalign{
{SU(3)_1\over SU(2)_2\times U(1)}\times SO(4)_1 &\equiv {SU(2)_2\over 
U(1)}\times U(1)\cr
 {SU(3)_2\over SU(2)_3\times U(1)}\times SO(4)_1 &\equiv \Big[{SU(2)_3\over
 U(1)}\times U(1)\Big]_{``D_6" } \ . \cr}}

Thus we see that the group is more intrinsically attached to $N=2$
theories or to their twisted, topological partner. This is very 
natural in view of the recent work of Dubrovin \Dub\ who has 
shown that in tft's a major role is played by the monodromy group
of a certain differential system. Under some assumptions, he was able
to prove that this group is generated by reflections. For the minimal
models  discussed in sect.2.4, these groups are the finite Coxeter
groups. More generally it is suggested that the group that I just 
exhibited is this monodromy group. Beside the cases of minimal
models quoted above, 
I have  checked that the group that occurs in the 
case $\CA^{(6)}$ is indeed  the monodromy group computed in \Dub. 
Also, the matrix $T$ of \IIIp\ receives a natural interpretation in 
that context, as the Stokes matrix between two asymptotic behaviors 
of some function. 

In contrast, if we return to the ordinary (``$N=0$") cft's or
integrable lattice models, it appears that 
the same group may be attached to two different models. These two different 
models, (pertaining for instance to $SU(2)$ and $SU(3)$, like in  
\IIIr), are in fact associated with two different {\it presentations}
of the group; the generators (resp. the roots) may,  depending on the case,
be organised into two or three sets of commuting (resp. orthogonal)
elements.  

As in the case of $SU(2)$ above, it is expected that non integer 
values of $G_{ab}$ describe certain subalgebras of the OPA. 
This is the case with the graph  $\CA^{(4)}$ with
$G_{ab}=2\cos{\pi\over h}$ which describes the subalgebra of the 
$\CA^{(h)}$ case generated by the left-right symmetric fields 
belonging to the orbit of the identity operator
under the $\IZ_3$  automorphism of the Weyl alcove, 
i.e. labelled by $\lambda=\bar\lambda
\in \{(1,1), (1,h-2), (h-2,1)\}$.

Finally it should be noted that all these considerations extend 
nicely to general $SU(N)$. This will be expounded in a separate 
publication. 



\newsec{Conclusion}
\nind 
I have shown that with each of the graphs introduced in the construction of
integrable lattice models one may associate in a natural way a Coxeter
group. Conversely given a Coxeter group with a certain presentation by
generators can one determine the graph? 
More precisely can one understand which class
of groups occurs in connection with, say, $SU(N)$? How does the 
$N$ colourability  of the graph manifest itself? Can one classify
the corresponding groups?

I have mentionned that within tft's, this group 
is the monodromy group introduced by Dubrovin. 
The most immediate question that arises is to find an interpretation
of this Coxeter group in the context of ordinary cft's and/or
lattice models. I have given hints that such an interpretation must 
be connected with the structure of the OPA, but at this stage 
this remains a vague and uncertain idea.

Clearly, there is ample room for more \dots reflections !

\bigskip\bigskip
{\bf Acknowledgements} I have benefited a lot from conversations with 
B. Dubrovin. It is also a pleasure to thank D. Bernard,
C. Itzykson, V. Pasquier and V. Petkova for useful discussions.

\listrefs

\bye